\documentclass[a4paper,11pt]{article}
\usepackage{pos}
\usepackage{wrapfig}

\title{High precision theory predictions in the Higgs sector}
\ShortTitle{High precision theory predictions in the Higgs sector}
\author[a]{Gudrun Heinrich}
\affiliation[a]{Karlsruhe Institute of Technology,\\
  Wolfgang-Gaede-Str. 1, 76131 Karlsruhe, Germany}

\emailAdd{gudrun.heinrich@kit.edu}

\abstract{We discuss recent progress in Standard Model predictions related to Higgs boson physics at the LHC and
 comment on  the combination of higher order corrections with potential effects of heavy New Physics parametrised by Effective Field Theories.}

\FullConference{%
  The Tenth Annual Conference on Large Hadron Collider Physics - LHCP2022\\
  16-20 May 2022\\
  online
}

\begin{document}
\maketitle

\section{Introduction}

Ten years after the discovery of the Higgs boson, there are impressive
achievements concerning the investigation of
its properties, for example the measurement of its mass with per-mille
accuracy, or the measurement of its couplings to vector bosons and quarks of the
third generation~\cite{ATLAS:2022vkf,CMS:2022dwd}.
Nonetheless, many open questions remain to be answered, for example whether the form
of the Higgs potential is as assumed in the Standard Model (SM), why the
masses of the fermions are so different, or where additional sources of
CP-violation could come from.
Questions ranging from the pattern of particle masses to cosmology are
related to the Higgs sector, therefore
the Higgs boson could serve as a guide towards possible answers.
As a consequence, a primary task of LHC physics consists in exploring
the many facets of the Higgs sector with highest possible precision,
in a joint effort of both experimentalists and theorists.

These proceedings pick some topics illustrating the recent progress in
high precision theory predictions in the Higgs sector.
More detailed reviews can be found e.g. in Refs.~\cite{Dawson:2022zbb,Huss:2022ful,Heinrich:2020ybq,LHCHiggsCrossSectionWorkingGroup:2016ypw}.

\section{Precision highlights in the Higgs sector}

\subsection{Higgs boson production in gluon fusion}

For the gluon fusion channel, impressive calculations have been
performed to decrease the theoretical uncertainties, both at fixed
order and beyond.
The uncertainty budget for the total cross section that has been put together a few years
ago~\cite{LHCHiggsCrossSectionWorkingGroup:2016ypw,Dulat:2018rbf} has
shrunk considerably since: NNLO corrections with full top-quark mass
dependence have become available~\cite{Czakon:2021yub,Czakon:2020vql,Davies:2019nhm}, the
uncertainties due to mixed QCD-electroweak corrections have been
reduced to about 0.6\%~\cite{Bonetti:2018ukf,Bonetti:2020hqh,Becchetti:2020wof,Becchetti:2021axs,Bonetti:2022lrk},
the threshold approximation has been overcome since quite some time
now~\cite{Mistlberger:2018etf},  the QCD corrections in the heavy
top limit (HTL) are available up to
N3LO~\cite{Chen:2021isd,Billis:2021ecs,Baglio:2022wzu}, even fully  differentially~\cite{Chen:2021isd},
and N4LO results exist in the soft-virtual approximation~\cite{Das:2020adl}.
Resummed N3LO+N3LL$^\prime$ results have been achieved for the Higgs-$p_T$ spectrum
and for the total cross section with fiducial cuts~\cite{Billis:2021ecs},
in Ref.~~\cite{Re:2021con} also including transverse recoil effects.
Publicly available programs such as {\tt n3loxs}~\cite{Baglio:2022wzu}, {\tt HTurbo}~\cite{Camarda:2022wti},
{\tt iHixs2}~\cite{Dulat:2018rbf}, {\tt ggHiggs}~\cite{Bonvini:2016frm} or {\tt  SusHi}~\cite{Harlander:2016hcx}  allow us to make detailed studies of theory uncertainties, as performed for example in Ref.~\cite{Baglio:2022wzu} for scale and PDF uncertainties.
A lesson learnt from the availability of N3LO results for Higgs- and vector boson production is that 7-point scale variations
at NNLO often do not capture the central value of the N3LO result, thus suggesting that global scale variations alone can be a too naive uncertainty estimate at the level of precision expected at such high perturbative orders.
A related problem consists in the fact that there is a mismatch between the perturbative order of the matrix elements (N3LO) and the PDF sets, which are not available at N3LO. For progress towards approximate N3LO PDFs see  Ref.~\cite{McGowan:2022nag}.

\subsection{Higgs boson production in association with a jet}

\begin{figure}[htb]
  \includegraphics[width=0.5\textwidth]{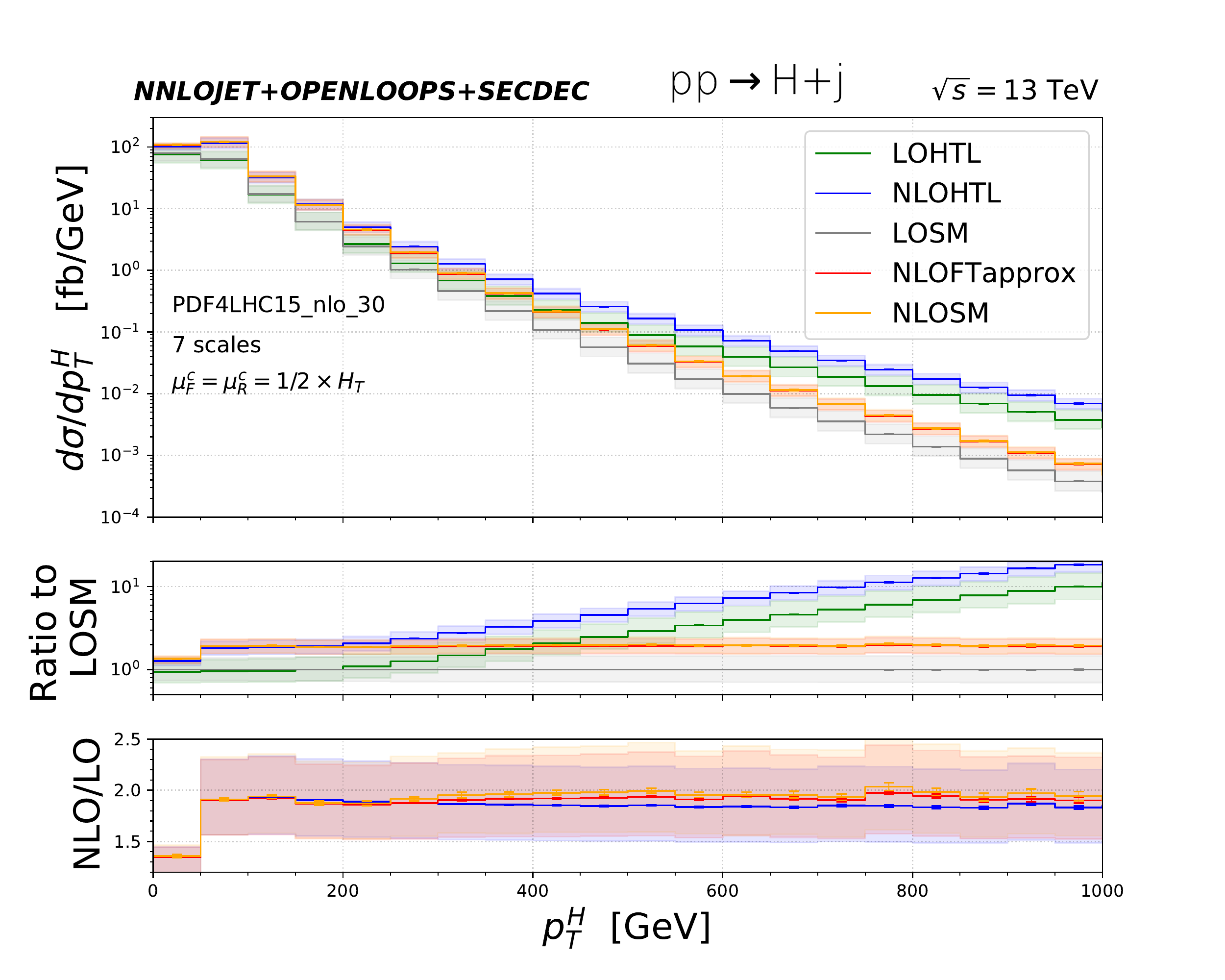}
  \includegraphics[width=0.47\textwidth]{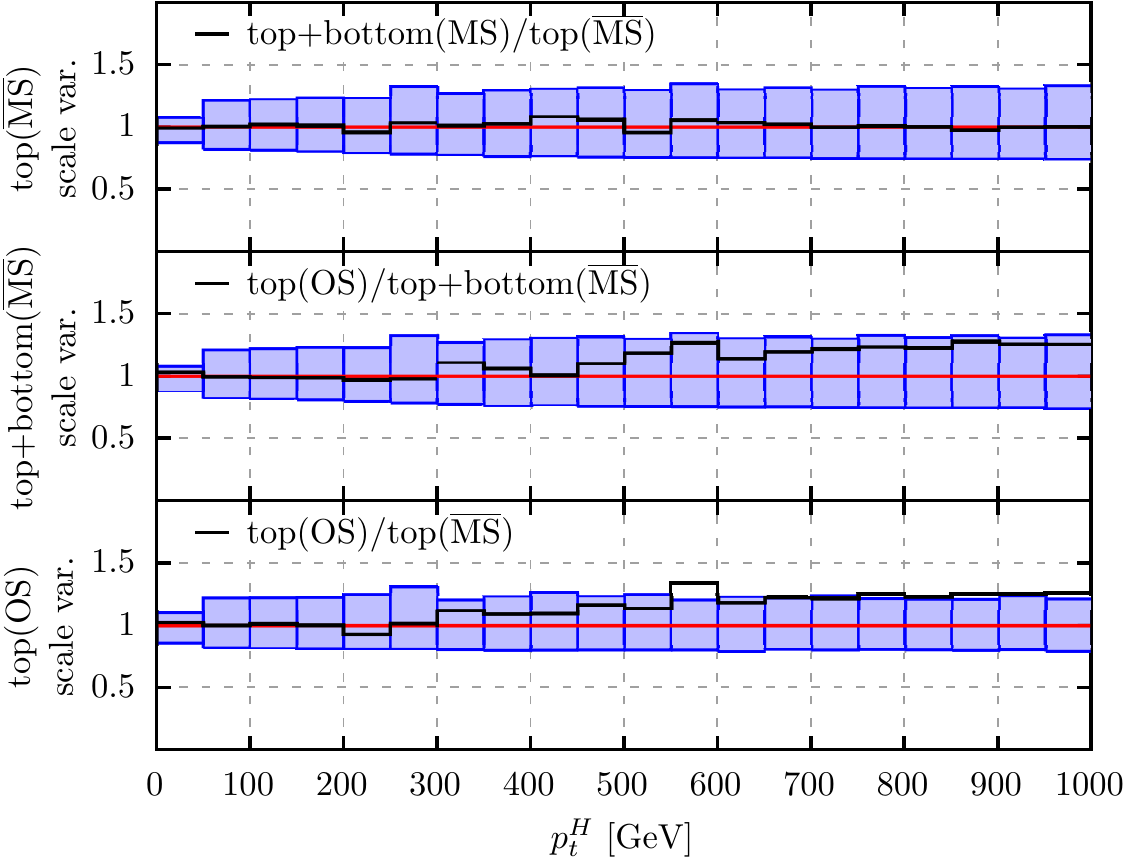}
\caption{Left: Transverse momentum distribution of the Higgs boson  in the
  full SM, in the heavy top limit (HTL) and in the ``approximate full
  theory'' (NLOFTapprox). The first ratio plot shows corrections relative to 
  LO SM, while the second ratio plot shows NLO results normalised to the respective LO prediction. Shaded bands denote scale uncertainties, error bars indicate integration uncertainties.
  Figure from Ref.~\cite{Chen:2021azt}.
Right: NLO/LO ratio for the transverse momentum distribution of the Higgs boson, with and without massive bottom loops and comparing on-shell and $\overline{\rm{MS}}$ schemes. Figure from Ref.~\cite{Bonciani:2022jmb}.}
\label{fig:hj_pth}
\end{figure}

The transverse momentum distribution of the Higgs boson is one of the most interesting observables to study at the LHC, for example due to its sensitivity to unknown particles circulating in the loop already at the leading order.
Heavy new particles would influence the tail of the $p_{T}^H$ distribution~\cite{Grazzini:2018eyk,Battaglia:2021nys}, therefore it is important to include the  quark mass dependence in the SM calculation for reliable predictions.
NLO predictions with full top-quark mass dependence have been calculated first in Refs.~\cite{Jones:2018hbb,Chen:2021azt}, with mass renormalisation in the on-shell scheme. Recently, results with both top- and bottom-quark mass dependence have been presented in Ref.~\cite{Bonciani:2022jmb}, see also Ref.~\cite{Lindert:2018iug} for earlier approximate results.
In Fig.~\ref{fig:hj_pth}\,(left), the approximations ``HTL'' (heavy top limit) and ``FTapprox'' (where only the 2-loop virtual corrections are in the heavy top limit) are compared to the full NLO calculation (NLOSM, orange). Fig.~\ref{fig:hj_pth}\,(right) shows the influence of including massive $b$-quark loops and the on-shell versus the $\overline{\rm{MS}}$ scheme on the ratio NLO/LO for the Higgs boson transverse momentum distribution.

%\subsection{Vector boson fusion}

\subsection{$ZH$ production}

\begin{figure}
  \centering
    \includegraphics[width=0.51\textwidth]{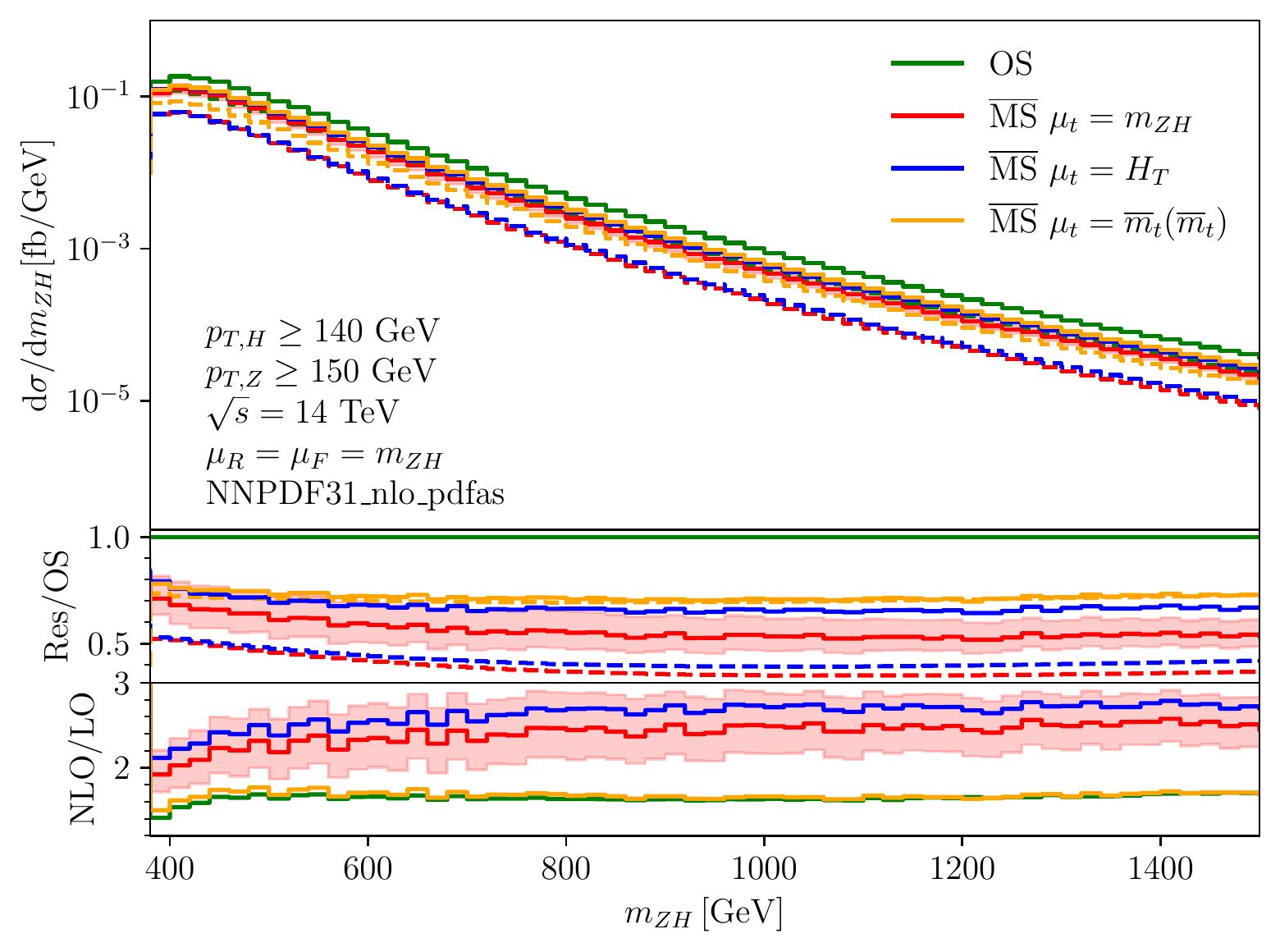}
    \includegraphics[width=0.43\textwidth]{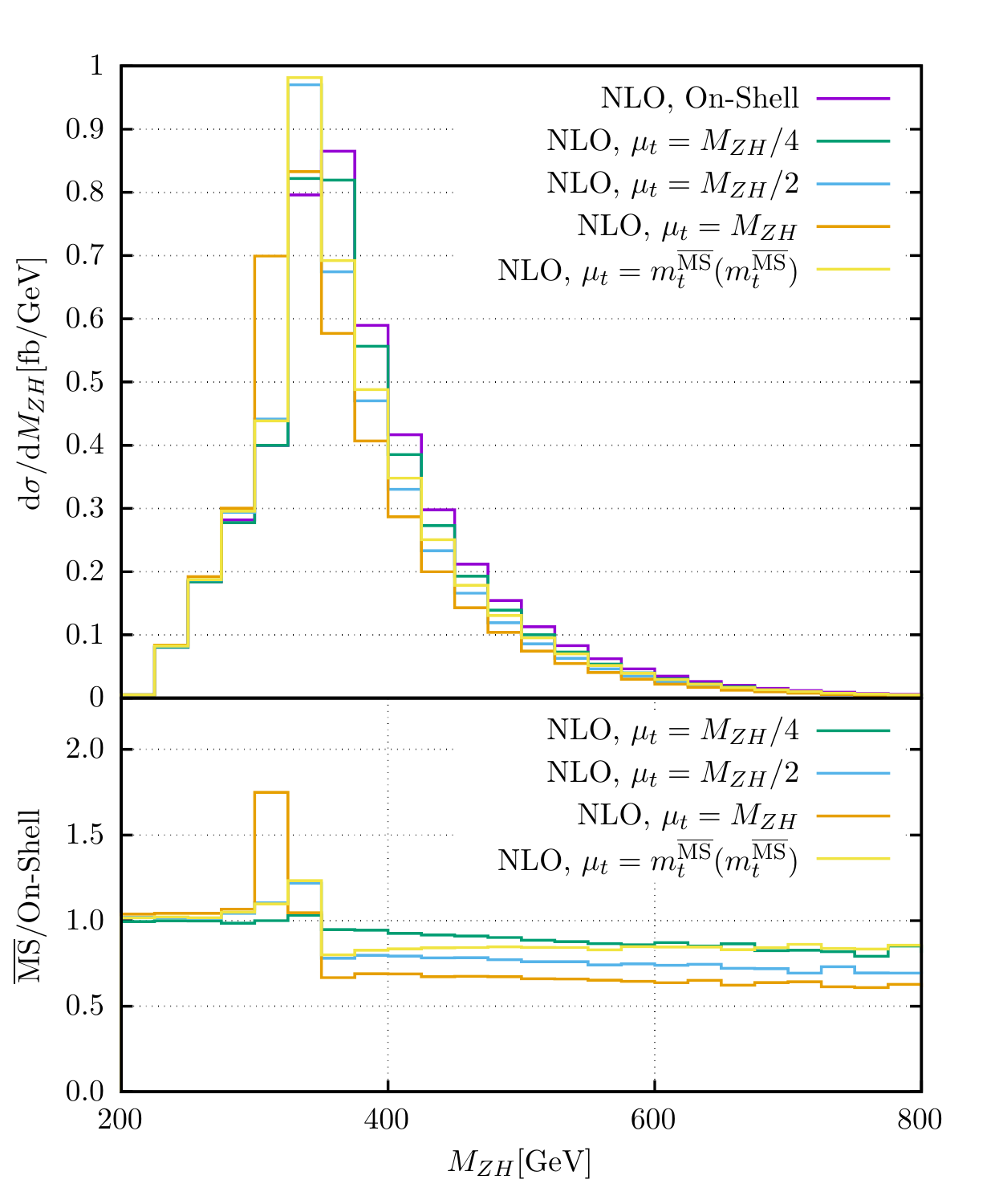}
  \caption{
  Left: Comparison of results at large values of the invariant mass $m_{ZH}$ using the on-shell
  or $\overline{\mathrm{MS}}$ scheme for the top quark mass  at LO (dashed) and NLO (solid), 
figure from Ref.~\cite{Chen:2022rua}. Right: Comparison in the range 200\,GeV$\leq m_{ZH}\leq $\,800\,GeV, figure from Ref.~\cite{Degrassi:2022mro}.}
  \label{fig:mt_scheme}
\end{figure}

Higgs boson production in association with a $Z$ boson is an
 interesting process as it probes both the Higgs boson
coupling to $Z$ bosons as well as to fermions.
Theory predictions for this process have been advanced in several respects recently.
The loop-induced gluon channel formally enters at NNLO with respect to the $pp\to ZH$ process,
and at its leading order accounts for about $6\%$ of the total cross section, thereby also introducing a relatively large scale uncertainty into the NNLO cross section. The  NLO corrections to the gluon channel increase the gluon-fusion cross section by about a factor of two, and reduce the scale dependence~\cite{Wang:2021rxu,Chen:2022rua,Degrassi:2022mro}.
In Refs.~\cite{Chen:2022rua,Degrassi:2022mro}, the dependence of the results on the top-quark mass renormalisation scheme at high energies has also been studied, as shown in Fig.~\ref{fig:mt_scheme} for the $m_{ZH}$ invariant mass, finding scheme uncertainties that exceed the scale uncertainties in the tail of the distribution.

Recent updates for $pp\to ZH$ production also include NNLO+PS results with $H\to b\bar{b}$, in the SM~\cite{Zanoli:2021iyp} as well as including anomalous Yukawa couplings~\cite{Haisch:2022nwz}. NNLO results with NNLO Higgs boson decays to massive $b$-quarks, including anomalous HVV couplings, have been calculated in Ref.~\cite{Bizon:2021rww}.
Results for $VH$+jet production at order $\alpha_s^3$ are also available~\cite{Gauld:2021ule}. The latter in addition represent a step towards a fully differential N3LO calculation of $pp\to VH$.

\section{Effective Field Theory and precision}

\begin{figure}
  \centering
    \includegraphics[width=0.5\textwidth]{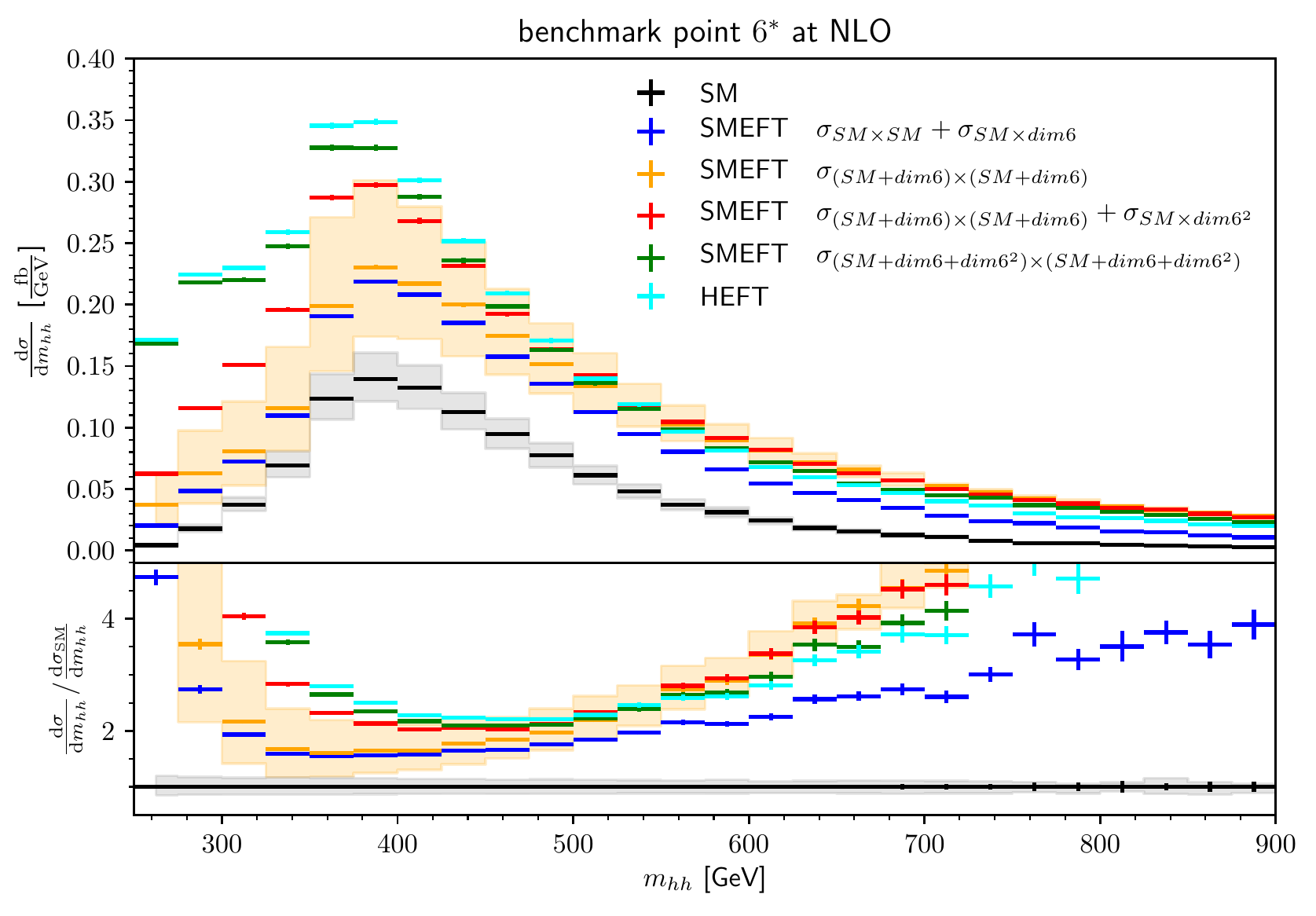}
     \includegraphics[width=0.46\textwidth]{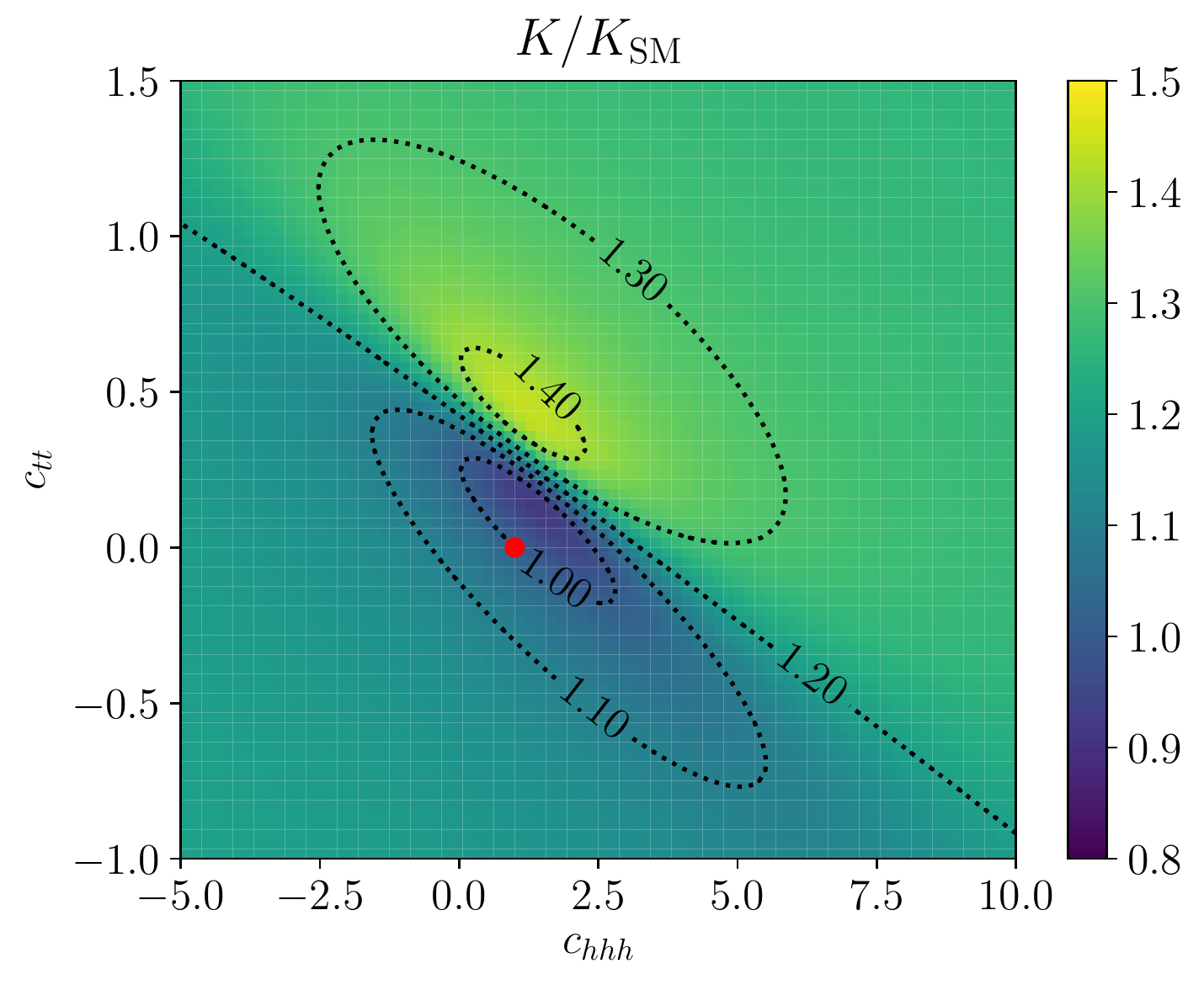}
   \caption{Left: Invariant mass distribution of the
   Higgs-boson pair for different truncation options in the SMEFT expansion at cross section level, for a benchmark point with $C_{H,\textrm{kin}}=0.56$,  $C_{H}=3.8$,  $C_{uH}=2.2$, $C_{HG}=0.0387$ at $\Lambda=1$\,TeV.
  Figure from Ref.~\cite{Heinrich:2022idm}. 
 Right: K-factor NNLO$^\prime$/LO relative to the SM K-factor in the $c_{tt}-c_{hhh}$ plane for Higgs-boson pair production, where $c_{hhh}$ denotes the trilinear Higgs self-coupling and $c_{tt}$ denotes an effective $t\bar{t}hh$ coupling. Figure from Ref.~\cite{deFlorian:2021azd}.}
  \label{fig:hh_smeft}
\end{figure}

The parametrisation of New Physics residing at higher energy scales in terms of Effective Field Theory (EFT) descriptions is a vast subject, treated in more detail elsewhere in these proceedings. The combination of higher order (mostly QCD) corrections with EFT expansions is important to extract reliable constraints on anomalous couplings from the data.
However, as EFTs are also expansions that need to be truncated at a certain order, additional uncertainties have to be considered.
For example, in Standard Model Effective Field Theory (SMEFT)~\cite{Buchmuller:1985jz,Grzadkowski:2010es,Brivio:2017vri}, differences between the inclusion of linearised dimension-6 terms at cross section level ($\sigma_\text{SM} + \sigma_{\text{SM}\times \rm{dim6}}$) or the square of amplitude-level dimension-6 terms ($\sigma_{\left(\text{SM}+\rm{dim6}\right)\times \left(\text{SM}+\rm{dim6}\right)}$) can be larger than the NLO QCD scale uncertainties. An example for the case of Higgs boson pair production is shown in Fig.~\ref{fig:hh_smeft}. 
Results of global fits also show differing patterns depending on the choice of the ``linearised'' or ``quadratic'' inclusion of dimension-6 operators, see e.g. Ref.~\cite{Ethier:2021bye} and related discussions~\cite{Brivio:2022pyi,Dawson:2021xei,Battaglia:2021nys,Trott:2021vqa,Martin:2021cvs}.
Another issue is the change of K-factors as a function of the anomalous couplings, an example is shown in Fig.\,\ref{fig:hh_smeft}\,(right).
Furthermore, it has been shown that the correlations between different Wilson coefficients can change considerably  with the energy scale if renormalisation group running effects are taken into account~\cite{Battaglia:2021nys,Banelli:2020iau}.

\section{Outlook}

Great progress has been achieved with regards to the availability of theoretical predictions for important observables in the Higgs sector at high orders in perturbative QCD, on the fixed order as well as on the resummation and parton shower side. In view of the shrinking scale uncertainties, other uncertainties such as mass effects, renormalisation scheme differences, electroweak corrections, PDF+$\alpha_s$-uncertainties and other non-perturbative uncertainties gain in relative importance and are relevant for the LHC precision program. Within EFT parametrisations of New Physics, the quest for precise predictions poses additional challenges, starting already with the choice of the EFT framework (HEFT or SMEFT), and requiring an assessment of truncation options and the validity range.
Nonetheless, making the step from ``anomalies'' in the data to established manifestations of physics beyond the Standard Model is in the cards of the LHC if we pursue the road of precision physics.

\subsection*{Acknowledgements}
This research was supported by the Deutsche Forschungsgemeinschaft
(DFG, German Research Foundation) under grant  396021762 - TRR 257.

%\begin{thebibliography}{99}

\bibliographystyle{JHEP}
\bibliography{Higgs_theory.bib}

%\end{thebibliography}

\end{document}